\input{aipcheck}

\documentclass[
 ]  {aipproc}

\layoutstyle{8x11single}
\def\gsim{\mathrel{\raise.3ex\hbox{$>$\kern-.75em\lower1ex\hbox{$\sim$}}}}
\def\lsim{\mathrel{\raise.3ex\hbox{$<$\kern-.75em\lower1ex\hbox{$\sim$}}}}

\begin{document}

\title{Free particle gauge invariance of the Maxwell equations.}

\classification{}
\keywords      {Maxwell equations, U(1) gauge, free electron, relativistic quantum equation, history pre-1900 physics }

\author{J.F. Geurdes\footnote{han.geurdes@gmail.com}}{
%%address={IFPA, D\'epartement AGO, Universit\'e de  Li\`ege, B-4000 Li\`ege, Belgium}
address={C. vd Lijnstraat 164, 2593 NN Den Haag, Netherlands}
}

\begin{abstract}
In a previous study \cite{Geurdes:1995}, it was demonstrated that Dirac's relativistic quantum equation for free electrons (DRQM) 
can be obtained from Maxwell's classical electromagnetic field equations (MaxEq). 
This raises fundamental issues about the demarcation between the classical and quantum domain and about spin. In
the present study it is assumed, however, that the derivation is physically relevant to cases 
where it is difficult to distinguish between classical and quantum physics.
The possibility of operating gauges on electron level is studied in view of the effect on the larger physical scale which provides a means to give experimental support to the presented pre-1900 history challenge. 
\end{abstract}

\maketitle

%%%%%%%%%%%%%%%%%%%%%%%%%%%%%%%%%%%%%%%%%%%%%%%%%%%%%%%%%%%%%%%%%%%%%%%%%%%%%%%%%%%%%%
%% MAINMATTER
%%%%%%%%%%%%%%%%%%%%%%%%%%%%%%%%%%%%%%%%%%%%%%%%%%%%%%%%%%%%%%%%%%%%%%%%%%%%%%%%%%%%%%

\section{Introduction}
The reason for the present study is twofold. In the fist place the study aims to apply and develop the already obtained inclusion of DRQM in MaxEq. In the second place it aims to describe a physical situation in order to enable an observational test of the consequences of this mathematical inclusion. The physical class of e.m. systems, we think, to which the derivation of DRQM from MaxEq and the present specific gauge analysis can be applied to, can be compared to the well-known 'school' phenomenon from marine biology. We may think of physical systems where particles behave in swarms or schools that have a nearly macroscopic classical size, but, like with a school of fish, the collective behavior is like a single fish i.e. 'like a single particle'. It make sense in those cases to employ characteristics of the 'deeper lying' partcipants in the school (i.e. characteristics of a single fish) because they determine the behavior of the school. This idea also connects to soft particles whose behavior are a result of collective behavior of constituent particles. The basic conventional description of the relation between quantum and classical radiation can be found in e.g. \cite{Heitler:1970}. 

In the application of the formalism, the author postpones the answer to the question  where exactly  the border between classical and quantum lies and how to, mathematically, get there. He simply notes that DRQM is usually seen as describing quantum phenomena, while MaxEq usually is seen as a set of equations valid for classical size systems. The 'old story' is that when Planck's constant can be neglected one is in the classical domain, but we simply have not a clue about the influence of Planck's constant in the twilight zone. Moreover, in our present case, if in the DRQM the Planck constant is neglected, we do not obtain MaxEq.

Let us start the recapitulation of the analysis with the classical MaxEq for the electromagnetic field. Here, we use
$x=(x_1,x_2,x_3,x_4)$ and $x_4=ix_0, x_0=t$ and we suppose  natural units .
\begin{eqnarray}\label{1}
\begin{array}{cc}{{\nabla\times\vec{E}}=-{\partial\vec{B}\over\partial{t}},~~~~\nabla\cdotp\vec{E}=\rho} \\
{ } \\
{{\nabla\times\vec{B}}={\vec{j}+{\partial\vec{E}\over\partial{t}}},~~~~\nabla\cdotp\vec{B}=0}
\end{array}
\end{eqnarray}
The $\vec{E}$  is the electric field vector and $\vec{B}$ the magnetic field vector, while, $\vec{j}$ is the current vector and $\rho$ the charge density.
Moreover, the dielectric constant and the magnetic permeability are chosen 
to be unity and $\hbar=c=1$ while the factor $4\pi$ is included in the charge density.
For referential purposes it will pay to have the set of Maxwell equations written in the previous form. This set of equations can, for
$\vec{F}=\vec{E}+i\vec{B}$, be reformulated into the following compound format.
\begin{eqnarray}\label{2}
{{\nabla\times\vec{F}}=i({\partial\vec{F}\over\partial{t}}+\vec{j}~~  )},~~~{\nabla\cdot\vec{F}=\rho}
\end{eqnarray}
In the previous ground-laying study \cite{Geurdes:1995}, a solenoid set of equations was assumed. From this assumption the following Quantum 
Dirac-like equation
\begin{eqnarray}\label{3}
\gamma^{\mu}D_\mu T \gamma^{\nu}D_\nu \psi = 0
\end{eqnarray}
was obtained. In the already cited work, the connection between $\psi$ and the expression in   (2) is given. Details can be found there.
Note that the usual summation convention over greek indices is employed, running from 1 to 4. Note also that the free-electron DRQM equation can be written 
like $\gamma^{\alpha}D_\alpha \psi = 0$   \cite{Thaller:1988}. Moreover, 
$T=\gamma^1\gamma^2+\gamma^2\gamma^3+\gamma^3\gamma^1$, with, $\gamma^\mu$ matrices obeying a Clifford algebra  
and $D_\mu=\partial_\mu - G_\mu$. The $G_\mu$, $\mu = 1,2,3,4$, represent the  deeper lying~  gauge functions operating on the free electrons. In order not to
confuse this set of gauge functions with the usual gauge functions for the MaxEq the notation,
$G_\mu$ is employed for  deeper lying  transformation. When the four vector $\psi(x)$, 
is gauge transformed, like $\psi~'(x)=exp(iR(x))\psi(x)$ a new set of equations, similar in form to   (3) is obtained.
\begin{eqnarray}\label{4}
\gamma^{\mu}D~'_\mu T \gamma^{\nu}D~'_\nu \psi~' = 0
\end{eqnarray}
Here, the $G_\mu$ transform as $G~'_\mu(x)=G_\mu(x) - i \partial_\mu R(x)$ hence, $D~'_\mu = \partial_\mu - G~'_\mu (x)$. In the present study, let us
turn the attention to the result of employing $\psi~'(x)=exp(iR(x))\psi(x)$ to the set of MaxEq. The $\psi \rightarrow \psi~' $ and
$G_\mu(x) \rightarrow G~'_\mu(x) $ are transformations at the level of 'a fish', the resultant set of Maxwell equations is at the level of  the school . 

We start with $\vec{F}=\vec{Q}(\vec{x})+\nabla\times\vec{C}$ and from those two vectors derive the transformed Maxwell equations. After the general 
$exp(iR)$ transformation the following set of irregular MaxEq is obtained.
\begin{eqnarray}\label{5}
\begin{array}{cc}{{\nabla\times\vec{E~'}} + {\partial\vec{B~'}\over\partial{t}}= -Im~\vec{j~'},~~~~\nabla\cdotp\vec{E~'}= Re \rho'} \\
{ } \\
{{\nabla\times\vec{B~'}} - {{\partial\vec{E~'}\over\partial{t}}}= Re~\vec{j~'},~~~~\nabla\cdotp\vec{B~'}= Im \rho' }
\end{array}
\end{eqnarray}
The primed entities result from the transformation of the four vector $\psi$ from   (3). We have
\begin{eqnarray}\label{6}
\vec{F'}=e^{iR}[\vec{F}+\phi \nabla R]
\end{eqnarray}
for the transformation of $\vec{F} \rightarrow \vec{F'}$ from   (2) where $\phi$ is a connection function. Secondly, we have
\begin{eqnarray}\label{7}
\rho' = \nabla\cdot[e^{iR}(\vec{Q}+i\phi \nabla R - i (\nabla R) \times \vec{C})]= \nabla\cdot\vec{Q~'}
\end{eqnarray}
for the $\rho \rightarrow \rho' $ transformation, with $\phi$ again the connection. Finally, the $\vec{j~'}$ as
\begin{eqnarray}\label{8}
\vec{j~'} = - {\partial\vec{Q~'}\over\partial t} + \nabla \times [e^{iR}(\vec{J}--i\vec{C}{\partial{R}\over\partial t})]
\end{eqnarray}
for the $\vec{j} \rightarrow \vec{j~'}$ transformation. Note that $\vec{j} = \nabla\times\vec{J}$ ~is supposed in the solenoid. The vectors, $\vec{Q}$
$\vec{C}$ and $\vec{J}$ can be quite generally chosen. Later in the paper more specific choices will be presented.

The irregular sytem of Maxwell equations from   (5) resembles
Maxwell's equations for Dirac monopoles \cite{Mikhailov:1993}. If a subsequent transformation in the magnetic vector is performed, 
resulting in a double primed system $\vec{B~''} = \vec{B~'} - Im \vec{Q~'}$ together with a transformation in the electric vector,
$\vec{E~''} = \vec{E~'} + Im[e^{iR}(\vec{J} - i \vec{C} {\partial R\over\partial t})]$ a regular, double primed, set of MaxEq like   (1) results. In particular we have
\begin{eqnarray}\label{9}
\rho~''= Re \rho~'  + \nabla\cdot Im [e^{iR}(\vec{J}-i\vec{C}{\partial{R}\over\partial{t}})]
\end{eqnarray}
for the transformed charge density and
\begin{eqnarray}\label{10}
\vec{j~''} = Re \vec{j~'} - \nabla\times(Im\vec{Q~'})-{\partial\over\partial{t}}Im[e^{iR}(\vec{J}-i\vec{C}{\partial{R}\over\partial{t}})]
\end{eqnarray}
for the transformed current density vector. The two latter equations will be the basis of subsequent analysis. Physically two types of transformations are 
active in the 'school'. First a  deeper lying  gauge that temporary may result in a monopole-like disbalance i.e. an irregular MaxEq. Secondly, a transformation of the MaxEq itself to
return to regularity. Note also that if we agree that $\vec{j~''} = \rho~''\vec{v}$
with $\vec{v}$ the velocity vector (\cite{Heitler:1970}), we must have,
\begin{eqnarray}\label{11}
\vec{v} = {{ Re \vec{j~'} - \nabla\times(Im\vec{Q~'})-{\partial\over\partial{t}}Im[e^{iR}(\vec{J}-i\vec{C}{\partial{R}\over\partial{t}})]}\over{Re \rho~'  + \nabla\cdot Im [e^{iR}(\vec{J}-i\vec{C}{\partial{R}\over\partial{t}})]}}
\end{eqnarray}
and according to special relativity $ ||\vec{v}~|| \leq 1 $, in the present units.

\section{A specific transformation}
In order to study a possible physical expression of the abstract transformations presented above, let us
take $\vec{E}=\vec{Q}(\vec{x})$. Because, $\vec{B}$ is a real vector it then follows from the expression 
$\vec{F}=\vec{E}+i\vec{B}=\vec{Q}(\vec{x})+\nabla\times\vec{C}$ that $\vec{C}=i\vec{W}$ follows where $\vec{W}$ a real vector. Hence, having 
$\vec{j}=\nabla\times\vec{J}$, from the solenoid condition and $\vec{B}=\nabla\times\vec{W}$, it is easy to see that $\nabla\cdot\vec{B}=\nabla\cdot\vec{j}=0$. 
It follows from the MaxEq  (1), especially the equation with the magnetic curl, together with, $\vec{E}=\vec{Q}(\vec{x})$ that 
$\nabla\times\vec{B}=\vec{j}=\nabla\times\vec{J}$. Because, curl on grad vanishes, we then may write for general $u=u(\vec{x},t)$ that
\begin{eqnarray}\label{12}
\vec{J}(\vec{x},t)=\vec{B}(\vec{x},t) + \nabla{u(\vec{x},t)}.
\end{eqnarray}
From the equation with the electric curl in the set MaxEq of  (1), together with $\vec{B}=\nabla\times\vec{W}$  it follows that 
\begin{eqnarray}\label{13}
\nabla\times\vec{E}=-{\partial\over\partial{t}}~\nabla\times\vec{W}.
\end{eqnarray}
Hence, it is possible to write $\vec{W}=-t\vec{E}(\vec{x})$.

Subsequently, let us study a simple U(1) gauge which only dependents on time $R=\omega_0~ t$. Because, 
$\nabla{R}=\vec{0}$ it then follows from   (7) above
that $\rho' = \nabla\cdot e^{iR}\vec{Q}~~$. Substitution in the previous   (8) then produces the following expression for $\rho~''$
\begin{eqnarray}\label{14}
\rho~''=Re~e^{i t\omega_0}~\nabla\cdot\vec{Q} + \nabla\cdot Im[e^{i t\omega_0}(\vec{B}(\vec{x},t) + \nabla{u(\vec{x},t)} -t\omega_0\vec{E}(\vec{x}))]
\end{eqnarray}
Rewriting the previous equation, using $\nabla\cdotp\vec{E(\vec{x})}=\rho(\vec{x})$ and $\nabla\cdotp\vec{B(\vec{x},t)}=0$ gives
\begin{eqnarray}\label{15}
\rho~''=[\cos(t\omega_0)-(t\omega_0)\sin(t\omega_0)]\rho(\vec{x})
\end{eqnarray}
when $\nabla^2 u(\vec{x},t)=0$. For the latter reason, in the present analysis, $\nabla u(\vec{x},t)$ is suppressed. From 
the conservation ${\partial\rho{~''}\over\partial{t}}~+~\nabla\cdot\vec{j~''}=0$, differentiating   (15), the transformed current density vector arises.
\begin{eqnarray}\label{16}
\vec{j}~''=\omega_0~[\omega_0~t\cos(t\omega_0)+2\sin(t\omega_0)]\vec{E}(\vec{x})
\end{eqnarray}
which is a specific form of   (8).

\section{Collective phenomena}
Upon taking a closer look at   (15), it appears that when the time at which a $exp(i t \omega_0)$ physically sets in (arbitrarily at t=0), then in the 
temporal development of the e.m. field, the transformed electric charge will dissapear. This will happen when $t=t_*$ and $t_*$ is the solution of 
$\cos(t\omega_0)-(t\omega_0)\sin(t\omega_0) =0$. Numerical approximation gives for the lowest $t_*>0$ that $t_*\approx {0.86\over\omega_0}$.
This suggests that at a certain moment in time because of gauge transformation of the underlying DRQM free particles the total charge in the 
electric field vanishes. The U(1) gauge $exp(it\omega_0)$ in the  underlying  makes the total charge to dissapear at $t=t_*$, without on the avergage to
stand still, because of $\vec{j~''} = \rho~''\vec{v}=\vec{0}$, when $\rho~''=0$ but not necessarily $\vec{v}=\vec{0}$. However substitution of $t=t_*$ shows
that $j~''$ is uneqal to  $\vec{0}$ when $\vec{E}(\vec{x})\neq\vec{0}$. Hence, $t=t_*$ is impossible for the $\exp(i t \omega_0)$ gauge in media that have a non vanishing charge density. If one starts with charge density equal to zero then the above analysis with a temporal gauge only,
is perhaps meaningless. However, in view of Sallhofer's analysis this starting-point can be usefull when discussing a Machian view \cite{Sallhofer:1986}. 
Because of the temporal development of the EM field, the onset of the deeper gauge is 
not possible prior to $t=t_*$. In fact, only for intervals, (k=1,2,3,...)
\begin{eqnarray}\label{17}
I_k=\{t\in {R} : t_*+(k-1)\pi<t<t_*+k\pi\}
\end{eqnarray}
can MaxEq transform in the $\exp(i t \omega_0)$ gauge. Hence, 'the individual fish in the school' cannot at all times 
perform all activities at its disposal because certain activities turn out to be impossible at the collective level.  If we take a look at equation (16),
then it is clear that $\vec{j~''}$ will also vanish for, $t=t_{**}$ a solution of $\omega_0~t\cos(t\omega_0)+2\sin(t\omega_0)=0$. From   (15) it follows
that for
$t=t_{**}$, the transformed density does not vanish, i.e. $\rho~''\neq 0$. Numerical approximation of $t=t_{**}$ shows that, $t_{**}\in I_1$ and is equal to
\begin{eqnarray}\label{18}
t_{**}\approx{2.289\over\omega_0}
\end{eqnarray}
Hence, at $t=t_{**}$ the average collective velocity vector vanishes, i.e. $\vec{v}=\vec{0}$. Two things catches the eye. Firstly this average zero velocity
caused by gauge transformation is repetitive. Hence, $\vec{v}=\vec{0}$ at $t=t_{**}+k\pi$. Secondly, by varying the $\omega_0$, the moment of $\vec{v}=\vec{0}$ 
can vary in time, although, $t\in I_k$ is a prerequisite in order to avoid contradiction.

\section{Description of a solution around $t=t_{**}$}
In order to gain insight into the behavior of MaxEq around $t=t_{**}$ let us in the first place assume that the development 
is such that 'around' $t=t_{**}$ during a certain time $\Delta{t}$, the current vector $\vec{j~''}$ vanishes while, generally, $\rho~''\neq 0$. Note
that during $\Delta{t}$ around $t=t_{**}$, we have $(\partial{\rho~''}/\partial{t})=0$ which is solenoid.   Let us also agree not to burden the
notation too much and therefore to suppress the double prime here. In this section Sallhofer's form of Dirac equation (\cite{Sallhofer:1986}) is employed. 

In Sallhofer's reformulation we will need the Pauli matrices (superscripts are indices here)
\begin{eqnarray}\label{18_1}
\sigma^1=\left ( \begin{array}{cc} 1 & 0 \\
                                                     0 &-1 \end{array} \right ),~
\sigma^2=\left ( \begin{array}{cc} 0 &-i \\
                                                     i & 0 \end{array} \right ),~
\sigma^3=\left ( \begin{array}{cc} 0 & 1 \\
                                                     1 &  0 \end{array} \right )
\end{eqnarray}
As is well-known or can be verified by hand, noting $\sigma^2\sigma^1=i\sigma^3$ and cyclic interchange, it is then possible to write the identity, ~ 
$i\vec{\sigma}\cdot\nabla\times\vec{Y}=( \vec{\sigma}\cdot\nabla)(\vec{\sigma}\cdot\vec{Y})~-~\nabla\cdot\vec{Y}$.   
This identity can, subsequently, be employed to rewrite the approximative solenoid MaxEq from (1) during $\Delta{t}$ around $t=t_{**}$, as
\begin{eqnarray}\label{18_2}
(\vec{\alpha}\cdot\nabla)\left ( \begin{array}{cc} \vec{\sigma}\cdot\vec{B} \\
                                               \vec{\sigma}\cdot\vec{E} \end{array} \right )+
                                      i\left ( \begin{array}{cc} 1_2 & 0_2 \\
                                                                            0_2 &-1_2 \end{array} \right ){\partial\over\partial{t}}
                                       \left ( \begin{array}{cc} \vec{\sigma}\cdot\vec{B} \\
                                               \vec{\sigma}\cdot\vec{E} \end{array} \right )=\rho~
                                       \left ( \begin{array}{cc} 1_2 \\
                                                                            0_2  \end{array} \right )=\rho~\hat{e}_1
\end{eqnarray}
Here, it is clear that
\begin{eqnarray}\label{18_3}
\vec{\alpha}=\left ( \begin{array}{cc}0 & \vec{\sigma} \\
                                               \vec{\sigma}&0 \end{array} \right ),~
\beta=\left ( \begin{array}{cc} 1_2 & 0_2 \\
                                                                            0_2 &-1_2 \end{array}\right )
\end{eqnarray}
and, for clarity, $1_2$ the $2\times 2$ identity matrix  $0_2$ the $2\times 2$, zeroes matrix. Furthermore, as usual in Dirac theory, we have
$\vec{\gamma}=i~\beta\vec{\alpha}$, while for short-hand:
\begin{eqnarray}\label{18_4}
\Psi=\left ( \begin{array}{cc} \vec{\sigma}\cdot\vec{B} \\
                                               \vec{\sigma}\cdot\vec{E} \end{array} \right ).
\end{eqnarray}
The 'Sallhofer' Dirac-like form that can be subsequently obtained from (20) is then
\begin{eqnarray}\label{18_5}
(\vec{\gamma}\cdot\nabla)\Psi~-~{\partial{\Psi}\over\partial{t}}=i~\rho~\hat{e}_1
\end{eqnarray}
Note that $\Psi$ is a $2\times4$ matrix, whereas 'our' $\psi$, from the previous sections, is $1\times4$. It is assumed here that $\psi$ and $\Psi$
refer to the same physical situation, which can be justified from $(\partial{\rho~''}/\partial{t})=0$ and $\vec{j~''}=\vec{0}$.    

During $\Delta{t}$ around $t=t_{**}$ it is assumed that $\Psi$=$\Psi(\vec{x},t)$, takes the form  
\begin{eqnarray}\label{18_6}
\Psi(\vec{x},t)=\Psi_0(\vec{x})\varphi(t)
\end{eqnarray}
with $\varphi(t)=\{-\eta t~+~\cos(t\omega_0)-(t\omega_0)\sin(t\omega_0)\}$. Now because, $t$ in $\Delta{t}$ around $t=t_{**}$ we may rewrite the partial
differential equation (23) as
\begin{eqnarray}\label{18_7}
(\vec{\gamma}\cdot\nabla)\Psi_0~+~\eta\Psi_0=i~\rho~\hat{e}_1
\end{eqnarray}
and $\eta$ such that $\varphi(t_{**})=1$ (resulting in $\eta\approx{-1.4774\omega_0}$ for the minimum $\eta$ ). The fact that $\varphi(t_{**})=1$ is approximated for  the whole temporal interval $\Delta{t}$ around $t=t_{**}$. Note that $\Psi_0$ is a 2x4 matrix. In this application of Sallhofer's reformulation of MaxEq it can be verified that the previous partial differential equation can be solved in the somewhat artificial case of $\rho_{\vec{k}} (\vec{x})=e^{i\vec{k}\cdot \vec{x}}$, with a solution matrix, $\Psi_{0,\vec{k}}(\vec{x})$. We have for singular solution,
\begin{eqnarray}\label{18_8}
\Psi_{0,\vec{k}~ Singl}(\vec{x})=A_{\vec{k}}~exp[-{\eta\over{3}}(\vec{\gamma }\cdot \vec{x})]{\hat{e}}_1
\end{eqnarray}
Varying the constant matrix $A_{\vec{k}}$, to, $A_{\vec{k}}(\vec{x})$ it is straightforward to obtain
\begin{eqnarray}\label{18_9}
A_{\vec{k}}(\vec{x})=(\eta + i~\vec{\gamma }\cdot \vec{k})^{-1}exp[({\eta\over{3}}\vec{\gamma }+i\vec{k})\cdot \vec{x})]
\end{eqnarray}
Note that for $\eta$ numerically small, we already have in numerical approximation $[1-{\eta\over{3}}(\vec{\gamma }\cdot \vec{x})][1+{\eta\over{3}}(\vec{\gamma }\cdot \vec{x})]=1+O(\eta ^2)$. Note also that the inverse matrix form can be written as
\begin{eqnarray}\label{18_10}
(\eta + i~\vec{\gamma }\cdot \vec{k})^{-1}=-\sum_{m=0}^{\infty }\eta ^{m} i^{~m+1} {(\vec{\gamma}\cdot \vec{k})^{m+1}\over{k^{2(m+1)}}} 
\end{eqnarray}
Moreover, it follows that $(\vec{\gamma}\cdot \vec{k})^{m+1}=k^{m+1}$, if, $m$ is an odd integer, while, $(\vec{\gamma}\cdot \vec{k})^{m+1}=k^{m+1}(\vec{\gamma}\cdot \vec{k})$ when $m$ is even. Subsequently, the consequences for the e.m. field solution, in terms of $(\vec{\sigma}\cdot \vec{B})$ and $(\vec{\sigma}\cdot \vec{E})$  will be demonstrated in $\Delta{t}$ around $t=t_{**}$. 

If one observes the form, ${\hat{e}}_1^{T}=(1_2,0_2)$, then it is clear that $(\vec{\sigma}\cdot \vec{B})$ is associated to odd $m$ terms in the sum. This will be explained subsequently.

From the artificial form $\rho_{\vec{k}} (\vec{x})=e^{i\vec{k}\cdot \vec{x}}$, a real distribution of charge can be derived with the use of Fourier analysis
\begin{eqnarray}\label{18_11}
\rho (\vec{x})=\int   e^{i\vec{k}\cdot \vec{x}}~ \tilde{\rho}({\vec{k}})~d^3\vec{k}
\end{eqnarray}
The Fourier transform $ \tilde{\rho}({\vec{k}})$ can also produce the associated solution of (25)
\begin{eqnarray}\label{18_12}
\Psi_0 (\vec{x})=\int  \Psi_{0,\vec{k}} (\vec{x}) ~ \tilde{\rho}({\vec{k}})~d^3\vec{k}
\end{eqnarray}
From (26), (27) and (28) for relatively small $\eta$ (30) gives
\begin{eqnarray}\label{18_13}
\Psi_0 (\vec{x})=- \sum_{m=0}^{\infty }\eta ^{m} i^{~m+1} \int  \tilde{\rho}({\vec{k}})~{(\vec{\gamma}\cdot \vec{k})^{m+1}\over{k^{2(m+1)}}}e^{i\vec{k}\cdot \vec{x}}d^3\vec{k}~{\hat{e}}_1
\end{eqnarray}
Hence, it follows that
\begin{eqnarray}\label{18_14}
(\vec{\sigma}\cdot \vec{B})=- \sum_{n=0}^{\infty }\eta ^{2n+1} i^{~2(n+1)} \int  \tilde{\rho}({\vec{k}})~{1\over{k^{2(n+1)}}}e^{i\vec{k}\cdot \vec{x}}d^3\vec{k}1_2
\end{eqnarray}
Because, from (22) it follows that 
\begin{eqnarray}\label{18_15}
(\vec{\sigma}\cdot\vec{B})=\left ( \begin{array}{cc} B_1 & B_3 - iB_2 \\
                                                                                 B_3 + iB_2 &-B_1 \end{array} \right )~
\end{eqnarray}
and because $1_2$ is diagonal, it follows that $\vec{B}$~vanishes in $\Delta{t}$ around $t=t_{**}$.   This entails, in turn, that the Fourier transform, $\tilde{\rho}({\vec{k}})$ of the, nonvanishing, charge density must ensure that  
\begin{eqnarray}\label{18_16}
 \sum_{n=0}^{\infty }\eta ^{2n+1} (-1)^{~(n+1)} \int  \tilde{\rho}({\vec{k}})~{1\over{k^{2(n+1)}}}e^{i\vec{k}\cdot \vec{x}}d^3\vec{k}=0
\end{eqnarray}

\section{Physical picture}
Hence, due to internal changes -mirrored in the gauge transformation in the Dirac equivalent equation- the motion of a collective of 'particles in an e.m. field', with a static density of charge, can come to a standstill in $\Delta{t}$ around $t=t_{**}$. The magnetic vector equally vanishes in $\Delta{t}$ around $t=t_{**}$. The Fourier transform of the density of charge must be such that (34) is satisfied because of the vanishing of the magnetic field vector. 

If, e.g. $\tilde{\rho}({\vec{k}})$, vanishes for relatively small $k=||\vec{k}||$ then the sum term in (34) can be approximated by
\begin{eqnarray}\label{18_17}
 \sum_{n=0}^{\infty }\eta ^{2n+1} (-1)^{~(n+1)}{1\over{k^{2(n+1)}}}\approx {-\eta\over{(k^2+{\eta}^2)}}
\end{eqnarray}
If, subsequently, $ \tilde{\rho}({\vec{k}})=(k^2+{\eta}^2) \sum_{n=1}^{N }e^{-i \vec{k}\cdot\vec{x}_n}$ there arises a nonvanishing charge density from
\begin{eqnarray}\label{18_18}
-\eta \int d^3\vec{k}~ e^{i(\vec{x}-\vec{x}_n)\cdot\vec{k})}=-8\eta{\pi}^3 \delta(\vec{x}-\vec{x}_n)
\end{eqnarray}
which implies that the area that can be reached by the system of 'charged particles in the e.m. field' in $\Delta{t}$ around $t=t_{**}$ does not contain point charges (represented by the position vector $\vec{x}_n$). Hence, the internal gauge transformation can be seen as a field internal process that drives charges out of the area that the system can reach in $\Delta{t}$ around $t=t_{**}$. The charge carrying particles will remain fixed at the boundary for  $\Delta{t}$ around $t=t_{**}$.  Note that because of $(\vec{\gamma}\cdot \vec{k})^{m+1}=k^{m+1}(\vec{\gamma}\cdot \vec{k})$, when $m$ is even, the electric field vector in general will not vanish in this area.

\section{Conclusion and discussion}
In the paper, a previous general discovery of the derivability of Dirac's relativistic quantum mechanical equation (DRQM) from Maxwell's electromagnetic 
field equations (MaxEq), is employed to picture a collective phenomenon based on the behavior of 'deeper lying' more elementary particles. The MaxEq equivalent DRQM reflects constituent quantum particles and can be compared to the behavior of a school of fish determined by the capacities of the single fish participating in the school. 
In this sense, it was possible to employ the simple elementary transformation $\exp{[it \omega_0]}$ that showed the surprising ability to let the collective field at $t\approx{2.289\over\omega_0} $
stop moving 'on the average'.  The periodicity is a consequence of the periodicity of the $\cot(\omega_0t)$ and $-2\tan(\omega_0t)$  functions from (15) and (16).
The scale $\omega_0$ multiplier
in the gauge determines the moment in time when this average zero velocity will occur. The application of the fundamental gauge $\exp(i t \omega_0)$,
producing the periodicity, is restricted to certain temporal intervals $t_*+(k-1)\pi<t<t_*+k\pi$ derived in the paper.  The restriction to the temporal intervals mirrors the fact that a single fish is also not free to do whatever he is capabale of doing because of limitations in the collective school.
Note that perhaps the collective $\vec{v}=\vec{0}$
bears relation to the famous measurement problem where a single quantum particle is believed to produce classical-size scale readings e.g.  
average $\vec{v}=\vec{0}$.

The question may be subsequently asked how such a simple gauge on the constitent particles (e.g. electrons) works in physics 
and how the scale $\omega_0$ is determined.   With further analysis using a different Dirac equivalence form, it was demonstrated that in $\Delta{t}$ around $t=t_{**}$ the magnetic field vector vanishes also and an example was given of a possible charge distribution. The gauge transformation representing a collective phenomenon can be translated to the physical possibility that charged particles are driven out of the area that the system can reach in $\Delta{t}$ around $t=t_{**}$.

Of course, the application  to collective phenomena does not answer the fundamental question of the derivation of DRQM from MaxEq as such (e.g.  \cite{Armour:2004}). 
However, it is not the first time that uninterpreted theory may lead to something interesting.
Moreover, the analysis weakly connects to the Machian view that particles are 'just' economic ways of describing reality (see also \cite{Sallhofer:1990}). At the same time, it could be 
statistical physics wisdom to consider both field and particle description of collective (meso) phenomena for after all the particle
wave duality is a real puzzle. Note that in a Machian sense fields can be 'just another economic way' to describe reality. It also has to be admitted that in the present paper, the underlying 
free-particle view is no doubt too ideal for real physical collectives. 
However, ideal collective classical forms similar to e.g. ideal gasses are expected to provide 
some approximative insight into physical nature. 

To end, the paper claims to study a genuine physically possible phenomenon lying 'on the border' between classical and quantum theory. It is claimed that studying systems of this size is relevant to e.g. measurement theory but has a right of its own also. In addition, the inclusion of DRQM in MaxEq makes sense in that twilight zone and is not in need of 'vanishing Planck constant' reasoning. The influence of the size of the constant in this twilight zone is altogether unknown. The obtained solution contains observational aspects about the magnetic field vector, the distribution of charges in relation to the gauge transformation of DRQM and the propagation velocity in time of the e.m. system under study. Finally, the Machian viewpoint can be helpfull in describing physical systems at the border of quantum and classical size, because we have precious little knowledge about the nature of this border. In fact in the discussion of the influence of the size of Planck's constant, it appears as though researchers only acknowledge classical v.s. quantal without 'something in between'. 

The possible physical borderline phenomenon found here can, qualitatively, be described as a void bubble with charges at its perimeter, that stands still and has a vanishing magnetic field vector and also has, most probably, a periodic re-occurence. Finding this in  (mesoscopic) physical reality will establish a physical validity to the 'challenge from pre-1900 history of physics'  \cite{Geurdes:1995}. 

%%%%%%%%%%%%%%%%%%%%%%%%%%%%%%%%%%%%%%%%%%%%%%%%%%%%%%%%%%%%%%%%%%%%%%%%%%%%%%%%%%%%%%%%%%%%%%
%% BACKMATTER
%%%%%%%%%%%%%%%%%%%%%%%%%%%%%%%%%%%%%%%%%%%%%%%%%%%%%%%%%%%%%%%%%%%%%%%%%%%%%%%%%%%%%%%%%%%%%%


%%\section {References}
\bibliographystyle{aipproc}   %% if natbib is available
\bibliographystyle{aipprocl} %% if natbib is missing

\bibliography{PlanetLatexHan}

\IfFileExists{\jobname.bbl}{}
 {\typeout{}
  \typeout{******************************************}
  \typeout{** Please run "bibtex \jobname" to optain}
  \typeout{** the bibliography and then re-run LaTeX}
  \typeout{** twice to fix the references!}
  \typeout{******************************************}
  \typeout{}
 }

\end{document}